\documentclass[preprintnumbers,amsmath,amssymb,prl,twocolumn]{revtex4}
\usepackage{dcolumn}
\usepackage{amsmath}
\usepackage{amssymb}
\usepackage{bm}
\usepackage{xspace}

\begin{document}

\title{Reply to Comment LCK1019 by Prokof'ev and Svistunov} 
\maketitle
\noindent
The purpose of Ref. \cite{Hove:2000b} was to study the geometry of real vortex loops 
(VLs) in $3D$ superconductors ($3DSC$) and superfluids ($3DSF$), and its connection 
to criticality. $3DSC$ are dual to $3DSF$, and vice versa \cite{Hove:2000b}. VLs 
originate with transverse fluctuations in the phase $\theta$ of the complex pairing 
field. A fundamental topological constraint is that at every point in a $3DSC$ or 
$3DSF$, one must have ${\bf \nabla} \cdot {\bf n} =0$; ${\bf n}$ is the vorticity 
defined by ${\bf \nabla} \times ({\bf \nabla} \theta) = 2 \pi {\bf n}$. {\it Vortices 
in $3DSC/3DSF$ cannot start or end inside the system}.\\ 
We related the Hausdorff dimension $D_H$ of the VLs to the
   anomalous scaling dimension $\eta$ of corresponding field theory.
   In the case of VL, this field theory is the {\it dual} of the
   original theory. The relation is $\eta + D_H =
   2$, where $\eta$ is defined via the correlation function of the
   field theory for the VL (i.e. the dual), $G({\bf x},{\bf y}) =
   \langle \phi^*({\bf x}) \phi({\bf y}) \rangle$.
At criticality, it takes the form  $G({\bf x},{\bf y}) = 1/|{\bf x}-{\bf y}|^{d-2+\eta}$ 
in $d$ dimensions and yields the probability of connecting two points ${\bf x}$ and 
${\bf y}$ with any directed continuous vortex path \cite{Hove:2000b}. {\it It is 
implicit \cite{Hove:2000b} that any path entering in computing  $G({\bf x},{\bf y})$ 
must be part of a (closed) VL.} Thus, 
$G({\bf x},{\bf y})= N_L({\bf x},{\bf y})/N_L$ where $N_L({\bf x},{\bf y})$ is 
the number of VLs passing through both  ${\bf x}$ and ${\bf y}$, while $N_L$ is 
the total number of VLs. A useful quantity is the corresponding probability 
$P_n({\bf x},{\bf y})$ of connecting ${\bf x}$ and ${\bf y}$ with a connected 
vortex path of length $n$, which must also be part of a closed VL. Hence, we have 
$P_n({\bf x},{\bf y})=N_L({\bf x},{\bf y};n)/N_L$, where $N_L({\bf x},{\bf y};n)$ 
is the number of VLs in the system passing through both  ${\bf x}$ and ${\bf y}$, 
when these two points are separated by paths of length $n$. Clearly,  
$G({\bf x},{\bf y}) =  \sum_n P_n({\bf x},{\bf y})$. We used the scaling form 
$P_n({\bf x},{\bf y}) = F(|{\bf x}-{\bf y}|/n^{1/D_H})/n^{d/D_H}$ \cite{Hove:2000b}, 
and assumed $F(0) \neq 0$, in order to have a finite probability of returning to 
the starting point by going around the entire closed VL. (Had we for instance assumed 
a form $F(t) \sim t^{\vartheta}; t \ll 1, \vartheta > 0$, we would have reached the 
erroneous conclusion that the probability of returning to the starting point by  
following a strictly closed VL path would be zero.) Moreover, the  distribution function 
$D(n)$ of closed loops of length $n$ is given by $D(n) = (1/V) \sum_{\bf x} P_n({\bf x},{\bf x})$.  
At the critical point, $D(n) \sim n^{-\alpha}$, with $\alpha = d/D_H+1$ \cite{Hove:2000b}. \\
The graphs of a high-temperature (HT) expansion of the {\it partition function Z} of the 
$3DXY$ model are closed paths that can be identified mathematically with sterically 
interacting closed loops \cite{Stan1971,PS2001,JS2005}. The graphs in the HT expansion 
of the {\it two-point correlation function G} are open-ended paths (OEP) \cite{PS2001}. 
Thus, the geometry of graphs contributing to Z differs from that of the graphs 
contributing to G \cite{PS2001}. The HT-graphs contributing to Z and G can in any case 
not be {\it physically} identified with VLs. \\
Ref. \cite{PS2005} computes $D_H$ of OEPs originating with a HT graph expansion of 
$\langle \phi^*({\bf x}) \phi({\bf y}) \rangle$ for a $|\phi|^4$-theory 
\cite{PS2005,PS2006_Private}, which is in the same universality class as the $3DXY$ model. 
This entails studying an analog of the above correlation function, namely 
\cite{PS2001,PS2005} 
$G({\bf x},{\bf y}) =  \sum_n P_n({\bf x},{\bf y})= \sum_n z_n({\bf x},{\bf y})/z_n$,
where $z_n = \sum_{{\bf y}}z_n({\bf x},{\bf y})$ is the number of open-ended paths
starting in ${\bf x}$ and ending up in {\it any} endpoint ${\bf y}$, while 
$z_n({\bf x},{\bf y})$ is the number of graphs of length $n$ starting in ${\bf x}$
and ending in ${\bf y}$. Importantly, $z_n$ clearly grows with the number of steps $n$. 
We write $z_n \sim n^{\vartheta/D_H}$, and 
$P_n({\bf x},{\bf y}) = F(|{\bf x}-{\bf y}|/n^{1/D_H})/n^{d/D_H}$ \cite{PS2005,JS2005}. 
In Refs. \cite{PS2005,JS2005}, one effectively defines a loop distribution function 
$D(n) \sim n^{-\alpha}$ even for these OEPs, by setting $|{\bf x}-{\bf y}| = a$ in $P_n$;
$a$ is a lattice constant. \\
Again $\alpha = d/D_H+1$ \cite{JS2005}, but the factor  $z_n$  in $P_n({\bf x},{\bf y})$ 
now requires, for consistency, that $F(t) \sim t^{\vartheta}$.  Including $\vartheta$, 
the scaling relation proposed in \cite{Hove:2000b} is modified as follows: 
$\eta + D_H = 2 - \vartheta$, where  {\it $\vartheta$ determines the asymptotic 
number $z_n$ of open-ended paths of length $n$ at the critical point in a system 
of OEPs \cite{JS2005}}. Computing $D_H$ of such OEPs,  Ref. \cite{PS2005} reports 
$D_H=1.7655(20)$. Using $\eta=0.038(4)$ (appropriate for the 
field theory of closed steric loops) \cite{Campostrini}, they infer 
$\theta = 0.1925(29)$ \cite{PS2005}.\\
While the definition of closed paths used in Ref. \cite{PS2005} is appropriate for 
OEPs, it is not for VLs.  OEPs are clearly less compact than VL's, whence their 
$D_H$ is smaller. Moreover, OEPs are unphysical objects in the context of VLs in 
$3DSC$, as is correctly implied on pg. 3 of Ref. \cite{PS2001}. Hence, computing 
$D_H$ for OEPs and relating it to $\eta$ of the dual field theory for VLs in a 
$3DSC$ \cite{PS2005,JS2005}, provides no test of the relation $\eta+D_H=2$. In 
$3DSC$ and $3DSF$, $\vartheta$ is fixed to zero by the topological constraint 
${\bf \nabla} \cdot {\bf n} =0$.\\
J. Hove and A. Sudb{\o}\\
Department of Physics, 
Norwegian University of Science and Technology, N-7491 Trondheim, Norway


\end{document}